\documentclass[10pt,a4paper,twocolumn,amsmath,amssymb,aps,prl,superscriptaddress]{revtex4-1}
\usepackage[latin9]{inputenc}
\usepackage{amsmath}
\usepackage{amssymb}
\usepackage{graphicx}
\usepackage{color}

\makeatletter

\pdfpageheight\paperheight
\pdfpagewidth\paperwidth

\makeatother

\begin{document}

\title{Multifractal metal in a disordered Josephson Junction Array}

\author{M. Pino}

\affiliation{Instituto de Física Fundamental, IFF-CSIC, Calle Serrano 113 b, Madrid
E-28006, Spain.}

\author{V.~E.~Kravtsov}

\affiliation{Abdus Salam International Center for Theoretical Physics, Strada
Costiera 11, 34151 Trieste, Italy}

\affiliation{L. D. Landau Institute for Theoretical Physics, Chernogolovka, Russia}

\author{B.~L.~Altshuler}

\affiliation{Physics Department, Columbia University, 538 West 120th Street, New
York, New York 10027, USA }

\author{L.~B.~Ioffe}

\affiliation{LPTHE - CNRS - UPMC, 4 place Jussieu Paris, 75252, France}

\affiliation{National Research University Higher School of Economics, Moscow,
Russia}
\begin{abstract}
We report the results of the numerical study of the non-dissipative
quantum Josephson junction chain with the focus on the statistics of many-body wave functions and local energy spectra. The disorder
in this chain is due to the random offset charges. This chain is one
of the simplest physical systems to study many-body localization.
We show that  the system may exhibit three distinct
regimes: insulating, characterized by the full localization of many-body wavefunctions, fully delocalized (metallic) one characterized
by the wavefunctions that take all the available phase volume and the
intermediate regime in which the volume taken by the wavefunction
scales as a non-trivial power of the full Hilbert space volume. In
the intermediate, non-ergodic regime the Thouless conductance (generalized
to many-body problem) does not change as a function of the chain length
indicating a failure of the conventional single-parameter scaling theory
of localization transition. The local spectra in this regime display
the fractal structure in the energy space which is related with the fractal structure of wave functions in the Hilbert space. A simple theory of fractality of local spectra is proposed and a new scaling relationship between fractal dimensions in the Hilbert and energy space is suggested and numerically tested.
\end{abstract}

\pacs{74.81.Fa, 71.27.+a, 74.40.Kb, 85.25.Cp}
\maketitle

\section{Introduction.}

The concept of single-particle localization introduced by Anderson
in 1958 \cite{An1958} was in fact prompted by the experiments of
Fehrer \cite{Feher1959} that studied electron spin relaxation of
P dopants in Si, a typical many-body problem. Despite its conceptual
importance, the many-body localization remained out of limelight until
the paper \cite{Basko2006} that proved the existence of disorder
driven transition in many-body systems. In contrast to the single
body localization, the properties of localization in the Fock space
of many-body system remain controversial. In particular, it is very
well established that single-particle localization in three dimensional
space happens as a result of a single transition. Only at the transition
point the properties of a single-particle wavefunction are described
by the scaling laws with anomalous dimensions \cite{Evers2008}. Recently
it was proposed\ \cite{Pino15} that this simple picture does not
hold for many-body localization: the many-body wavefunction retains
anomalous dimensions in a finite parameter region. In this region,
the volume occupied by a typical wavefunction scales as anomalous
power, $D$, of the full Hilbert space volume that continuously changes
from $D=0$ in the insulator to $D=1$ in a fully delocalized state.
In a qualitative agreement several groups have found that the dynamics
in this region is often described by non-trivial power laws that are
neither diffusive nor localized\ \cite{Lev2014,Luitz2016,BarLev2017,Santos2015}.

The anomalous dimension, $0<D<1$, of the wavefunction implies that
a many-body system does not visit all allowed configurational space
in the course of time evolution, i.e. non-ergodicity. Qualitatively,
the non-ergodic behavior is very natural in strongly disordered quasiclassical
systems where strong disorder prevents the system from visiting all
Hilbert space while the quasiclassical parameter makes localization
very difficult. Empirically such behavior is well known for spin glasses
with large spin that break ergodicity without full localization. The
possibility of a delocalized non-ergodic behavior is very important
for the interpretation of the data on atomic systems such as \cite{Schreiber2015,Bloch2017}
because it implies that slow dynamics does not mean full localization.
The non-ergodic state of the superconducting systems can be detected
by the noise measurements that is expected to show strong violation
of FDT\cite{Pino15}; in line with these expectations a giant noise
was reported recently close to superconductor-insulator transition \cite{Tamir2016}.
A more detailed discussion of the physical properties in this regime
can be found in \cite{Pino15,Biroli2017}.

The existence of a non-ergodic regime gets additional support from
the results\ \cite{Altshuler2014,Al2016,Biroli2010} for the single-body localization on Caylee tree and random regular graphs, the problems
that are believed \cite{Al1997,Basko2006} to be similar to the many-body localization. Even though there is no doubt that single-particle
localization on Caylee tree displays the non-ergodic behavior, the
applicability of this result to many-body problems and even to random
regular graphs was questioned recently \cite{Tikhonov2016,Garcia2016_1,Metz2017}.
Unlike the single-particle problem on the Caylee tree the full many
body localization does not allow analytical treatment; the numerical
analysis remains inconclusive for available system sizes, its results
allow interpretation as in terms of ergodic Griffiths phase \cite{Argarwal2015,Gopalakrishnan2016}
as well as the fractal non-ergodic state \cite{Serbyn2016,Garcia2016_1,Deng106,Serbyn2016_2}.
The ambiguity is partly due to the fact that the non-ergodic regime
appears in a narrow range of parameters in the studied models.

In this paper we report the evidence for the appearance of a
non-ergodic regime in the model where this regime is expected to appear
in a wide range of parameters. Qualitatively, one expects that this
situation is realized in the systems with large quasiclassical parameter
in which the localization is driven by another parameter that can
be changed independently.

\begin{figure}[t!]
\begin{centering}
\includegraphics[width=0.8\columnwidth]{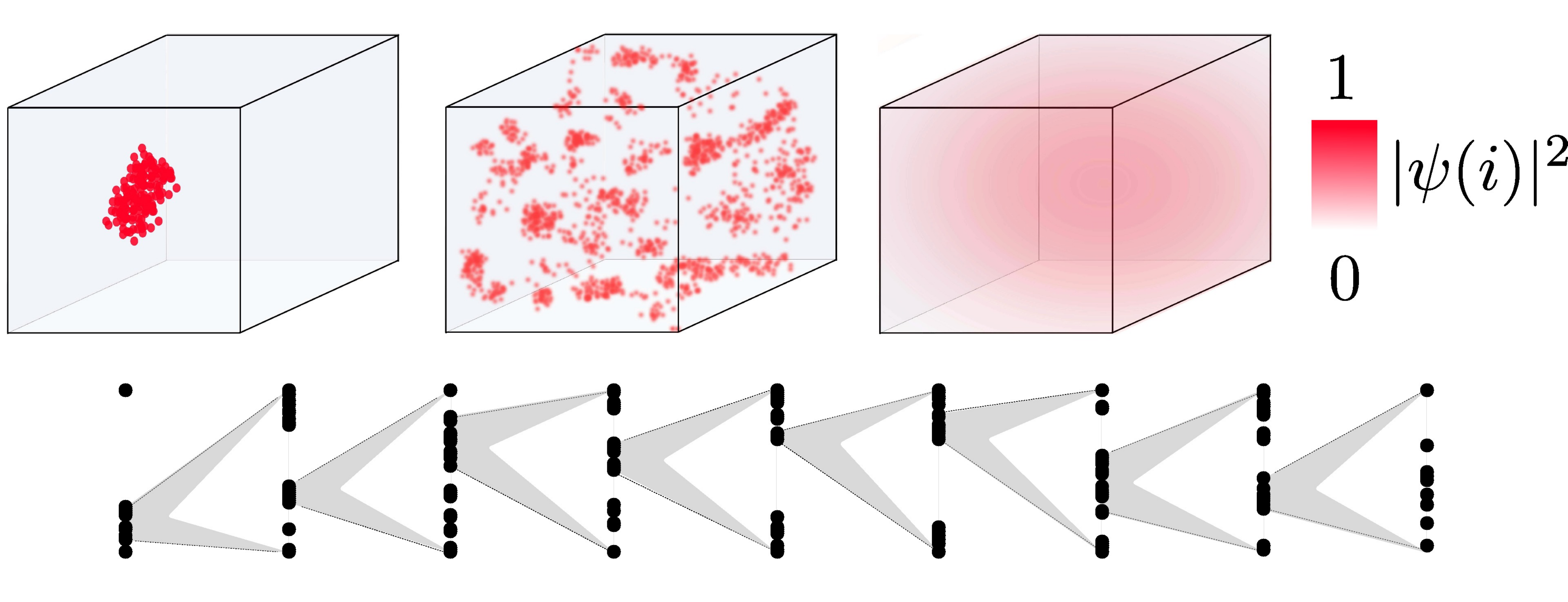}
\par\end{centering}
\caption{Upper panel - cartoon of a many-body wavefunction in three distinct
regimes localized (left), non-ergodic metallic and ergodic states
(right). These regimes differ by the ratio of the total number of
Fock states, $N$, and the support set $\Omega$, where wavefunction
is significant. In a localized state, the volume of the support set,
$\Omega$ is finite, or at most logarithmic, so $\Omega/N<\ln(N)/N$.
For the non-ergodic metal the support set forms a fractal structure,
so $\Omega/N<N^{-\nu}$ with $\nu<1$. In the ergodic phase, the support
set scales with the dimension of the Many-Body space, so $\Omega/N\sim1$
and the probability is uniform. Low panel - the local spectrum (energy
levels for which wavefunction is significant on a given site) is
similar to the random Cantor set. The full spectrum (left) contains
5000 energy levels forming groups separated by large gaps. Zoom into
each group produces similar structure at all energy scales. \label{fig1}}
\end{figure}

The wavefunction in the non-ergodic state can be visualized as hybridization
of distant resonances that happen to be very close in energy, see
Fig. \ref{fig1} In contrast to a single
body problem, in the many-body one the number of states grows exponentially
with the order of the perturbation theory that makes it likely to
find weakly coupled very distant but strongly mixed resonances. The
states formed by the linear combinations of these resonances form
a \emph{mini-band} that is responsible for delocalization. All energy
scales in this mini-band are small and determine the Thouless energy,
$E_{Th}$, for the whole system that might become much smaller than average
level spacing so that the effective Thouless conductance $g=E_{Th}/\delta\ll1$
is small and size independent in a wide parameter range. Our numerical results
confirm this qualitative picture.

The formation of mini-bands characterized by a small Thouless energy
can be viewed as a consequence of weak interaction strength which
is nevertheless sufficient for delocalization. This unusual regime
is known to occur in critical power-law banded matrices with parametrically
small off-diagonal elements $\left\langle H_{nm}^{2}\right\rangle =b^{2}/(n-m)^{2}$
with $b\ll1$\ \cite{Cuevas2007}. In this model the dimensionless
conductance turns out to be small $g\sim b$ and size independent.

The conductance that varies by orders of magnitude as a function of
parameters but remain size independent distinguishes the many-body
localization from localization in three dimension where $g$ is constant
only in  the critical region where $g\sim1.$ However, the difference disappears
in both localized and ergodic regimes where the conductance becomes
a fast function of the size.

 The structure of this paper is as follows. In section \ref{sec:hamiltonian} the model is introduced and
its physical realization as an array of Josephson junctions is explained. A  brief description of the numerical methods used in this paper can be found in section \ref{sec:numerical}. The theory of fractal local energy spectrum in a multifractal regime is presented in section \ref{sec:correlations}. In this section the correlation function $K(\omega)$ of the local densities of states is introduced and studied in a simple model of multifractality of many-body wave functions that generates a fractal local energy spectrum characterized by the fractal dimension $D_{s}$. A new scaling relationship between this fractal dimension and the fractal dimension $D_{2}$ of many-body wave functions in the Hilbert space is derived. The definition of the many-body Thouless energy and the Thouless conductance is also done in this section in terms of $K(\omega)$ and it is shown that multifractality leads to size-independent Thouless conductance.
This theory is tested by the numerical results for $K(\omega)$ in Sec.\ref{numerics-Kw}. In section \ref{sec:scaling} the many-body Thouless conductance is evaluated numerically and it is shown that it is size-independent in a wide range
of parameters of the model. The fractal dimensions $D_{1}$ and $D_{2}$ are numerically evaluated in section \ref{sec:fractaldimensions}. The new scaling relationship between the fractal dimensions in the Hilbert and energy space is tested in Sec.\ref{sec:relationship}. In section \ref{sec:statistic} the $r$-statistics of many-body energy levels is studied and an approximate position of the many-body localization transition in the parameter space is located. In Conclusion the main results of the paper are summarized.

\section{Model and experimental realization} \label{sec:hamiltonian}
A simple and physically realizable model is provided by the idealized Josephson junction chain with a high
ratio of Josephson, $E_{J}$ and charging energies $E_{C}$, $E_{J}\gg E_{C}$:
\begin{align}
H= & E_{J}\sum_{i=1}^{L}\cos\left(\phi_{i}-\phi_{i+1}\right)+E_{C}\sum_{i=1}^{L}\left(\hat{q}_{i}-n_{i}\right)^{2},\label{Eq:Ham}
\end{align}
where $\hat{q}$ is the operator conjugated to the phase $\phi_{i}$, $e\,n_{i}$ is the random static offset charge.
  We will set $E_C=1$ which fixes energy units in the following.
All the calculations below have been done for the closed loop, $|q_{L+1}\rangle=|q_{1}\rangle$. This geometry is experimentally relevant because it allows to protect the chain from the noise coming from dc lines (see below).

In this system the localization transition
is driven by temperature. Unexpectedly, the many-body wavefunction
becomes localized at \emph{high }temperatures $T\geq T_{MBL}$ : $T_{MBL}\sim E_{J}^{2}/E_{C}$\ \cite{Pino15}.
On the other hand, in the whole range $T\gg E_{J}$ the classical
dynamics of the phase is only weakly affected by the Josephson couplings
and is almost periodic indicating that the system is non-ergodic in
this regime. The low-temperature behavior of a related disordered system has been recently studied in the context of a Bose glass\ \cite{Vo15,Ce17}.
For numerical analysis reported here we have restricted
the allowed charging states by $-Q\leq q\leq Q$ with $Q=2$. We assume
that $n_{i}$ is distributed uniformly in the interval $(-W,W)$ and
focus on the regime of relatively strong disorder $W=10$. Note that
while in the realistic chain the offset charges $n_{i}$ are completely
random, their effective range is $-1/2\leq n\leq1/2$ because larger
$n$ can be eliminated by the shift of $q$. In the model with restricted
$-Q\leq q\leq Q$, this is not true and the range of $n$ becomes
relevant.

\begin{figure}[h!]
\includegraphics[width=0.8\columnwidth]{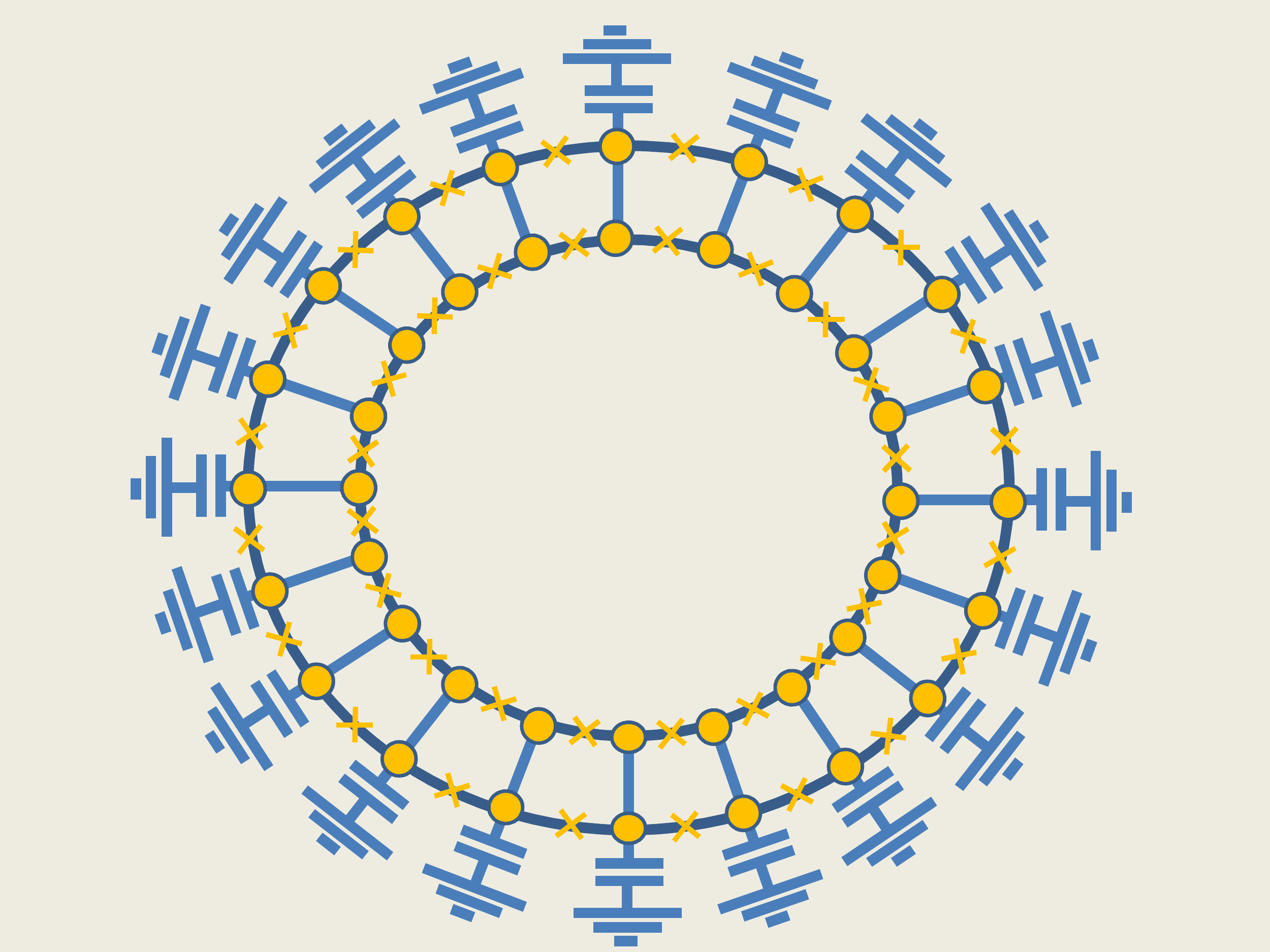} \caption{Schematic experimental Josephson-junction array setup in a form of a closed loop. \label{fig2}}
\end{figure}

The sketch of the experimental setup is shown in Fig.\ \ref{fig2}. First of all we note that in order to control $E_{J}$ one needs
to connect the superconducting islands by SQUID loops.
The closed geometry significantly reduces the noise coming from environment. Similar physics should hold in an open chain but in order to be decoupled from the environment the dc lines that lead to them should contain superinductance or other decouplers.

In order to ensure $E_{J}<\Delta$ and to neglect thermal quasiparticles we need low transparency junctions. These junctions do not have significant capacitance. It is important that they have small size and do not contain parasitic two-level systems.
In order to implement the model Hamiltonian Eq.(1)  the experimental setup should have large capacitance to the ground that would dominate ground capacitance. So, the correct setup should contain these capacitors as additional elements (shown in Fig.\ \ref{fig2}).
All these elements should have low loss (in particular low loss tangent of the ground capacitance implies that one should be careful with the choice of dielectric, better to avoid any dielectric in fact).

The "smoking gun" evidence of the non-ergodic extended phase is the enhanced   noise that by far exceeds the one predicted by the Fluctuation-Dissipation Theorem (FDT). Thus studying the noise and comparing it with the linear response at the same frequency one can detect the violation of FDT.
Assuming that the effective loss tangent (that takes into account the participation ratio)
can be kept at the level of $10^{-4}-10^{-5}$ we expect that one can ignore the dissipation at frequencies higher than 20 kHz. This sets the range for the frequency response $f\sim 10$ kHz.

The main idealization of our approach is the neglect of all excitations except those of the model Hamiltonian Eq.(1). Especially dangerous are the ones associated with the quasiparticles.
To avoid thermal quasiparticles we need $T, E_{J} <0.1\,\Delta$. Thus the realistic estimate of parameters of our model are  $E_{J}\sim 10-100\,mK$ and  $E_{C}\sim 1-10\,mK$.

An important issue is the non-equilibrium quasiparticles that are ubiquitous in the systems considered.
Note that the mere presence of the stationary quasiparticle in the island does do any harm.
The problem is the motion of quasiparticles between the islands that change the random offset charge.
The rate of this motion depends on the experimental setup, it can vary between $1$ kHz \cite{Wang2014} and minutes \cite{Bell2016}.
In any case it is much lower than the frequency at which the response (noise) should be studied. It can be viewed as a random change of the offset charge configuration, similar to numerical experiments in which we studied quantities averaged over many configurations. The effect of the non-thermal quasiparticles will be exactly to reproduce the averaging in numerical experiments.

\section{Numerical method}\label{sec:numerical}

We perform the exact diagonalization of the
restricted model (\ref{Eq:Ham}) and analyze a few states at energies
${E=E_{gs}+\bar{\epsilon}\,\mathfrak{W}}$, where $E_{gs}$ and $\mathfrak{W}$ are the ground-state energy and the many-body
band-width. The numerical diagonalization of Hamiltonian Eq.\ (\ref{Eq:Ham}) has been done
by two methods. In the first one, we have used partial diagonalization
to obtain a few eigenstates at a given energy density with ARAPCK's shift
invert mode\ \cite{Le98,Luitz15} . In the second one, we have used
a full diagonalization to obtain all the eigenstates. The former method
allows the computation of system with sizes up to $L=11$, while
the last one is only capable of solving sizes up to $L=8$.
We will mainly present results for eigenstates at energy $\bar{\epsilon}=0.1$.
Partial diagonalization of Many-Body system is more efficient away from the middle of the spectrum than at the band center, where the mean level spacing is much smaller.
Thus, the choice of energy density $\bar{\epsilon}=0.1$ allows to reach larger system sizes.

The number of disorder realization of Hamiltonian
Eq.\ref{Eq:Ham} used to average a given quantity has been chosen to make sensible
error bars. Error bars are computed as the standard deviation of the
population of measurements given by different realization of the disorder.
We notice that smaller values of $E_{J}$ requires larger number of
disorder realizations. Thus, for $E_{J}\leq4$, we have used around
$10^{4}$ realizations and for $E_{J}$=14 around $10^{3}$.

Note that at the largest system size $L=8$ attainable for full diagonalization the size of the Hilbert space was $N\sim 10^{6}$, so that together with the  number of disorder realizations $\sim 10^{4}$ and 10 different values of $E_{J}$ the computational cost was really enormous.

\section{LDoS correlation function and fractality of local energy spectrum.}\label{sec:correlations}

A central part of this work is to compute the {\it many-body Thouless energy} \cite{Kravtsov2015, Serbyn2016_2}.
To this end we employ the correlation function $K(\omega)$  of Local Densities of States (LDoS) between two points $E+\omega/2$ and $E-\omega/2$ in the energy space.
It is defined by \cite{Cuevas2007}:
\begin{equation}
K_{E}(\omega)=\frac{{N^{2}}\sum_{\alpha,\beta}\overline{|\psi_{\alpha}(i)|^{2}
|\psi_{\beta}(i)|^{2}\delta(E_{\alpha}-E_{-})\delta(E_{\beta}-E_{+})}}
{\sum_{\alpha,\beta}\overline{\delta(E_{\alpha}-E_{-})\delta(E_{\beta}-E_{+})}}
\label{eq:K(w)}
\end{equation}
where $N$ is the dimension of Hilbert space, $E_{\pm}=E\pm\omega/2$,
$\psi_{\alpha}(i)$ is the wavefunction at site $i$ in the Hilbert space
of charge quantum numbers, the bar means average over all different
charge states and disorder realizations.
The denominator in Eq.(\ref{eq:K(w)}) serves to factor out the effect of level repulsion at small $\omega$ and extract a pure correlation of different wave functions at a site. At larger $\omega$  the level repulsion can be neglected and   the factor $N^{2}/\sum_{\alpha,\beta}\overline{\delta(E_{\alpha}-E_{-})\delta(E_{\beta}-E_{+})}$
reduces to $\rho(E)^{-2}$, where $\rho(E)=N^{-1}\sum_{\alpha}\overline{\delta(E-E_{\alpha})}$ is the global density of states.

\begin{figure}[t]
\includegraphics[width=0.7\linewidth]{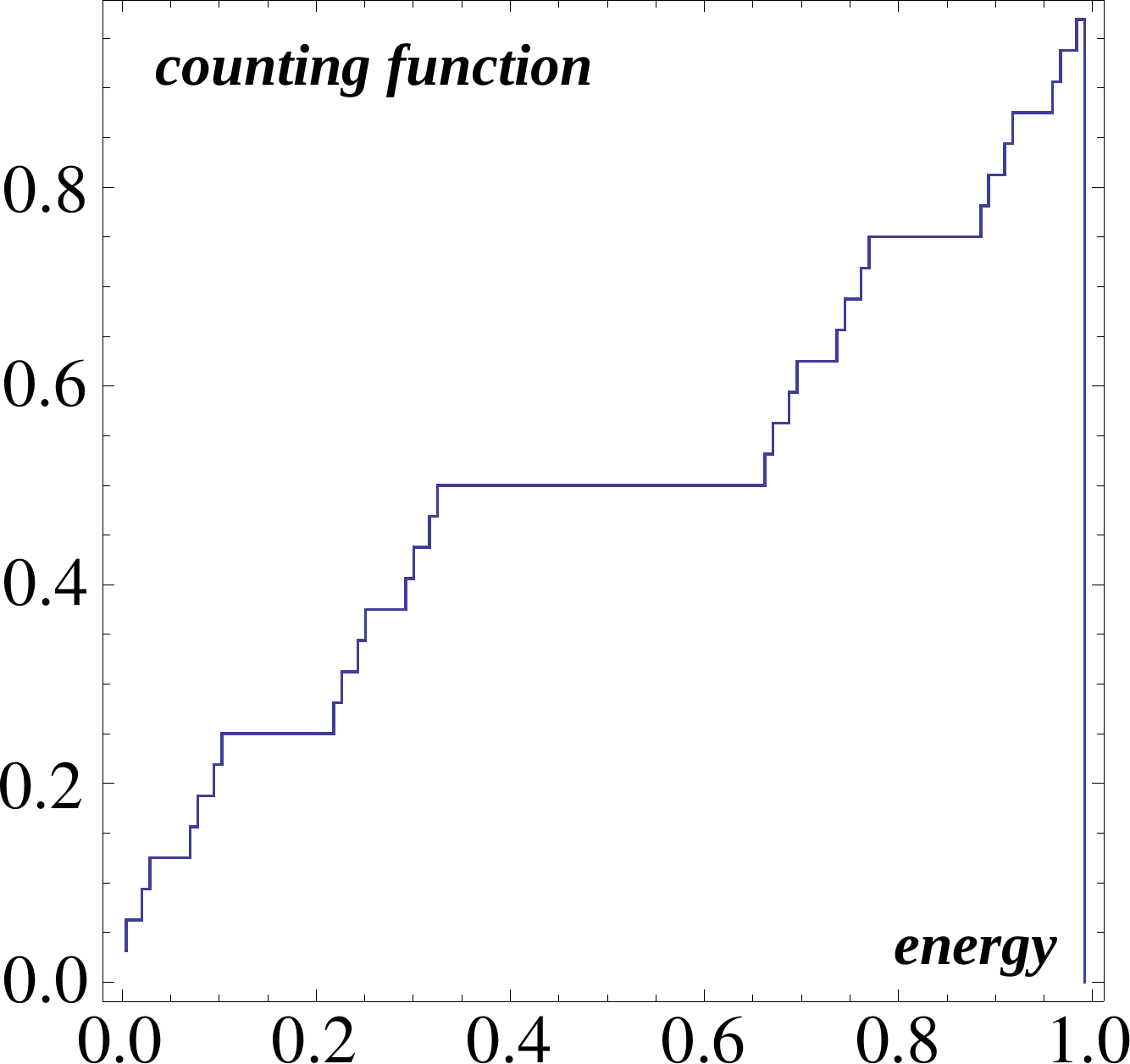} \caption{(Color online)  Counting function for the standard Cantor set with fractal dimension $D_{s}=\ln 2/\ln 3$. Each new level corresponds to a step in the vertical direction. The plateaus correspond to a gap in the spectrum. There is a middle gap of the width 1/3; in each of the side bands there is its own middle gap of the width 1/9, etc.  \label{Fig:Cantor} }
\end{figure}
\begin{figure}[t!]
\includegraphics[width=0.7\linewidth]{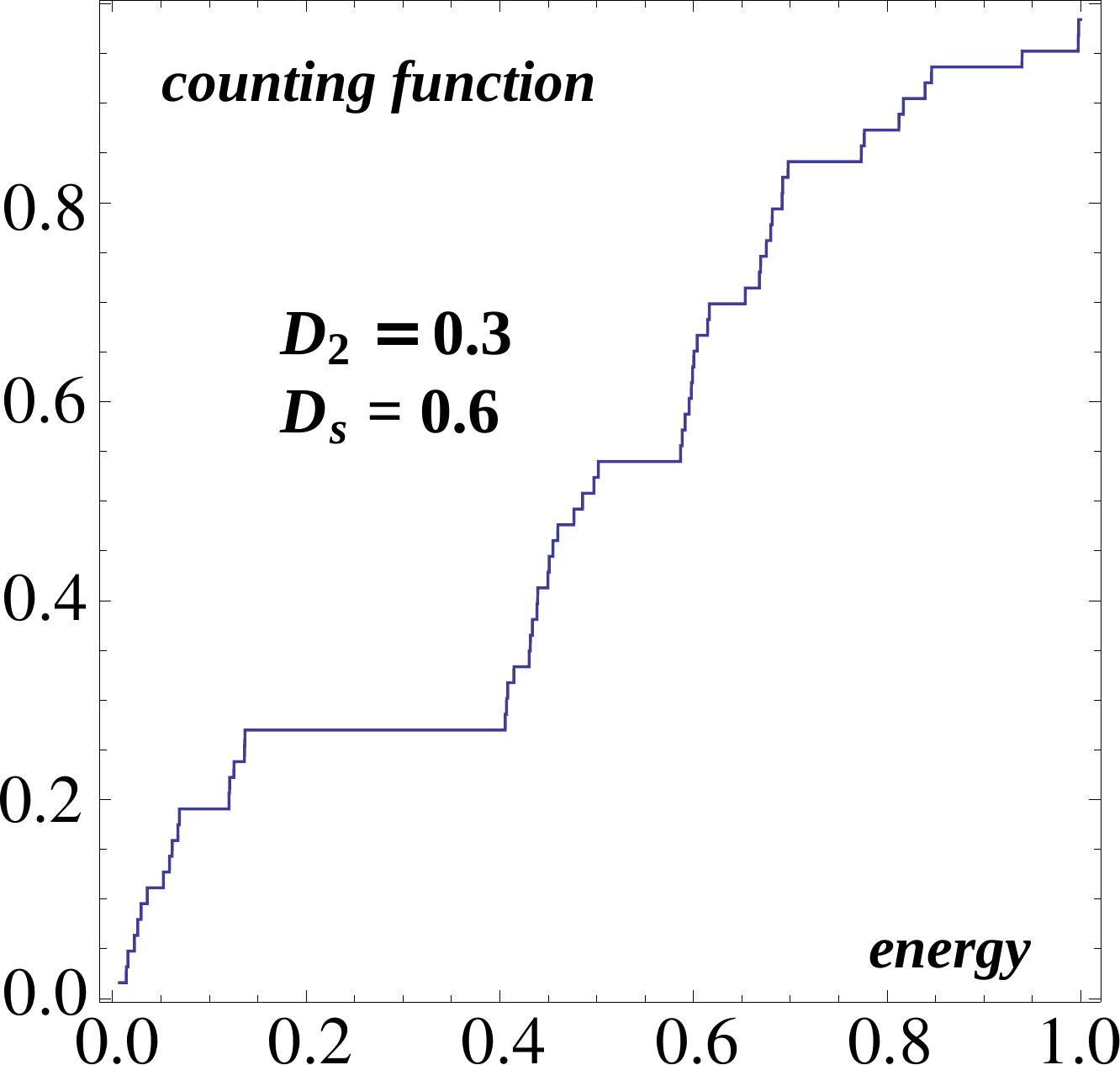} \caption{(Color online) Counting function for a random Cantor set, Eq.(\ref{K-num}) with $D_{2}=0.3$,  generated by statistically independent identically distributed level spacings
with the distribution given by Eq.(\ref{eq:P(Delta)_power}) at $D_{s}=0.6$.   \label{Fig:RandomCantor1.0} }
\end{figure}

For ergodic normalized wave function their overlap is perfect and energy-independent $|\psi_{\alpha}(i)|^{2}\,|\psi_{\beta}(i)|^{2}\sim N^{-2}$, and the correlation function $K(\omega)\sim 1$ is a constant. For localized wave functions it is exponentially small for most of disorder realizations but  in rare events   which happen with probability $\sim 1/N^{2}$ it is very large $\sim N^{2}$.

For the non-ergodic multifractal wave functions the correlation function $K_{E}(\omega)$
is a power-law in $|\omega|$ which low-energy cutoff is the Thouless energy. This power-law is a signature of fractality of local energy spectrum which can be illustrated by the following simplified model.

In this model we assume \cite{Cuevas2007} that wave functions are grouped
in certain families  sharing  the same fractal support set in the Hilbert space.   Each support set consists of $M\sim N^{D_{2}}$ sites, where $D_{2}$ is its Housedorff dimension of the support set.  Since multifractal wave functions are extended over the support set and have vanishingly small amplitude outside it, by normalization  their amplitude on a support set is   $  |\psi|^{2} \sim M^{-1} \sim N^{-D_{2}}$. Under this assumption at $\omega\gg \delta\sim (N\rho(E))^{-1}$   one can
represent the correlation function $K(\omega)=\int K_{E}(\omega)\,\rho(E)^{2}\,dE$ as follows:
\begin{equation}
K(\omega)= M^{-2}\sum_{a,b}\delta(\omega-E_{ab}),\label{K-num}
\end{equation}
where $E_{ab}=E_{a}-E_{b}$ is the difference between energies for the states belonging to the {\it same family} which support set includes the observation point $i$ where LDoS is evaluated.

Clearly, the distribution of the level spacings for such states ("local level" spacings) may differ qualitatively from the global level spacing distribution. Indeed, Eq.(\ref{K-num}) contains only those levels which states belong to the same family, other states are completely discriminated out. This may lead to large gaps between the local levels inside which levels of other families are situated.  These gaps are statistically much more probable than for the global spectrum where all levels are taken into account.

A natural assumption (which will be confirmed by our numerics) is that the fractality in the Hilbert space  corresponds to a fractality in the local energy spectrum. In other words, fractality is a property of eigenstates in the "Hilbert space-local energy spectrum" extension rather than only a spacial property of eigenstates.

A well-known example of a fractal spectrum is the standard Cantor set (see Fig.\ref{Fig:Cantor}). Remarkably, a similar hierarchical structure of gaps
can be generated (seeFig.\ref{Fig:RandomCantor1.0}) in the simple model of statistically independent local level spacings with the power-law probability density identical for all spacings  \cite{KravtsovScardicchio}:
\begin{equation}
P(\Delta) \sim \frac{(E_{{\rm Th}})^{D_{s}}}{\Delta^{1+D_{s}}},\,\,\Delta>E_{{\rm Th}},
\label{eq:P(Delta)_power}
\end{equation}
where  $E_{{\rm Th}}$ gives the low-energy cutoff.
The exponent $D_s$ is the measure of the fractality of the local energy spectrum, $0<D_s<1$.

One can easily calculate $K(\omega)$ in the  model  described by Eq.(\ref{K-num}), where $E_{ab}=\sum_{n=b}^{a}\Delta_{n}$, and each $\Delta_{n}$ is i.i.d. random variable with the power-law distribution Eq.(\ref{eq:P(Delta)_power}):
\begin{widetext}
\begin{equation}
k(t)  =M^{-2}\sum_{a,b}^{M}\langle e^{-it\sum_{n=b}^{a}\Delta_{n}}\rangle
=  M^{-1}\left(2\Re\frac{p(t)}{1-p(t)}+1\right)-2M^{-2}\,\Re\left[\left(1+ \frac{p(t) -
p(t)^{M }}{(1-p(t))}\right)\,\frac{p(t)}{1-p(t)}\right]\label{eq:k(t)}
\end{equation}
\end{widetext}
where we introduce  the Fourier transforms:
\begin{align*}
k(t) & =\int K(\omega)e^{-i\omega t}d\omega\\
p(t) & =\int P(\Delta)e^{-i\omega t}d\Delta.
\end{align*}
For any distribution function its Fourier transform $p(0)=1$, $\Re p(t\neq0)<1$
and $p(t\rightarrow\infty)=0.$ In addition, for the power-law distribution
function Eq.\ (\ref{eq:P(Delta)_power}) one obtains in the region of interest $E_{Th} t\ll 1$:
\begin{equation}\label{1-p}
1-p(t)\sim(t E_{Th})^{D_{s}}\ll 1,\;\;\;\;(E_{Th} t\ll 1).
\end{equation}
Thus in Eq.(\ref{eq:k(t)}) there is a small parameter $M^{-1}\ll 1$ and a large parameter $(1-p(t))^{-1}$. Competition between them leads to two different regimes.

If $M (1-p)\gg 1$, or $\omega\ll E_{{\rm Th}}\,N^{D_{2}/D_{s}}$, the term proportional to $M^{-2}$ in Eq.(\ref{eq:k(t)}) can be neglected.
Transforming
back to the frequency space we get the power law dependence for $N^{D_{2}/D_{s}}\,E_{{\rm Th}}\lesssim\omega\lesssim E_{{\rm Th}} $:
\begin{equation}
K(\omega)= 2M^{-1}\int \Re\frac{p(t)}{1-p(t)}\,e^{i\omega t}\,\frac{d t}{2\pi}\sim  \frac{(E_{{\rm Th}})^{-D_{s}}}{M\,\omega^{1-D_{s}}}.
 \label{Kw}
\end{equation}
In the opposite limit $M\,(1-p)\ll 1$ the leading in $1-p$ term in Eq.(\ref{eq:k(t)}) is $1-\frac{M}{3}\,(1-p)$. Thus  one obtains a faster decay of $K(\omega)$ for  $1\gtrsim\omega\gtrsim E_{{\rm Th}}\,N^{D_{2}/D_{s}}$:
\begin{equation}
K(\omega)= -(M/3)\int \Re(1-p(t))\,e^{i\omega t}\,\frac{d t}{2\pi} \sim\frac{M\,(E_{{\rm Th}})^{D_{s}}}{\omega^{1+D_{s}}}.
\end{equation}
Finally, at the smallest $\omega\lesssim E_{{\rm Th}}$ the correlation function $K_{E}(\omega)$, Eq.(\ref{eq:K(w)}), (as well as $K(\omega)\propto K_{E}(\omega)$) reaches the limit set by the inverse participation ratio $I_{2}(N)=\overline{\sum_{i}|\psi(i)|^{4}}\propto N^{-D_{2}}\sim 1/M $:
\begin{equation}\label{IPR-limit}
 K_{E}(\omega\rightarrow+0)=c\,N\, I_{2}(N) \sim N^{1-D_{2}}.
\end{equation}
The coefficient $c$ is not 1 because of the de Broglie oscillations of random wave functions. For completely random oscillations of {\it real} wave functions in the Wigner-Dyson Random Matrix Theory this coefficient is equal to 1/3.

In what follows we use Eq.(\ref{IPR-limit}) with $c=1/3$ to {\it define} the Thouless energy:
\begin{equation}\label{Thouless-def}
K_{E}(\omega=E_{{\rm Th}})= \frac{1}{3}\,N\,I_{2}(N).
\end{equation}
Next, we define the {Many-Body Thouless conductance}:
\begin{equation}\label{Thouless-cond}
g=\frac{E_{{\rm Th}}}{\rho(E)\,N},
\end{equation}
as the ratio of the Thouless energy and the many-body mean level spacing $\delta=(\rho(E)\,N)^{-1}$.

Comparing Eq.(\ref{Kw}) with Eq.(\ref{Thouless-def}), where $P_{2}(N)=N^{-D_{2}}=M^{-1}$, we conclude that in the non-ergodic multifractal phase the Thouless energy must be proportional to $N^{-1}$, and thus the Many-Body Thouless conductance $g$ should be independent of $N$ just as the conventional single-particle Thouless conductance at the critical point of the Anderson localization transition.

As we will see in the next section, this remarkable property is confirmed numerically in the broad interval of parameters of our model which corresponds to $g$ varying by almost two orders of magnitude as these parameters are changing.

Finally, assuming $E_{{\rm Th}}\propto N^{-1}$ and using $M=N^{D_{2}}$ we
find the relationships between the critical exponents that control $K(\omega)$:
\begin{equation}\label{def-beta-mu}
K(\omega)=\frac{A}{\omega^{\mu}}\sim\frac{N^{\beta}}{\omega^{\mu}}.
\end{equation}
For  $E_{{\rm Th}}\lesssim\omega\lesssim E_{{\rm Th}}\,N^{D_{2}/D_{s}}$ we obtain
from Eq.(\ref{Kw}):
\begin{eqnarray} 
\beta & =&D_{s}-D_{2},\label{betamu1} \\
\mu & =&1-D_{s}\label{betamu2}.
\end{eqnarray}

These equation should be compared with the ones for the critical point
of the 3D Anderson model, where the standard Chalker's scaling \cite{Chalker90} holds:
\begin{equation}
\mu=1-D_{2}.
\end{equation}
Thus the standard Chalker's scaling corresponds to   $D_{s}=D_{2}$ which implies that the fractality has the same
dimension in the Hilbert space (represented by "sites" $i$) and in the frequency space.

In general, for $\beta\neq0$ we have a generalized Chalker's scaling:
\begin{equation}\label{gen-Chalker-scaling}
\mu+\beta=1-D_{2}.
\end{equation}
In the next section we will show that the model considered in this paper corresponds to $\beta>0$ (see Fig. \ref{fig4}).

A special limiting case of  vanishing fractality of local energy spectrum corresponds to $D_{s}=1$ in Eq.(\ref{eq:P(Delta)_power}). In this case
$$1-p(t)\sim - C\;it\,  \ln(it\,E_{0}),$$ where $C$ and  $E_{0}$ are the pre-factor in front of the power-law and its low-$\Delta$ cutoff. In this case Eq.(\ref{Kw}) predicts a very slow, logarithmic decrease of the correlation function at $E_{0}\lesssim\omega\lesssim M\,C$:
\begin{equation}\label{Ds1}
K(\omega)\sim \frac{1}{M\,C\,\ln(\omega/E_{0})}.
\end{equation}
\begin{figure}[t]
\includegraphics[width=0.9\linewidth]{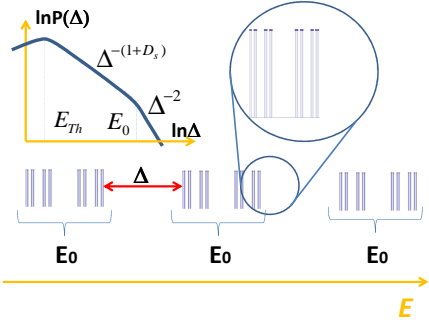} \caption{(Color online) Minibands in the local energy spectrum corresponding to Eqs.(\ref{Kw}),(\ref{HE-plateau}). $E_{0}$ sets the maximal scale of hierarchical structure. "Cantor sets"  of the width $\sim E_{0}$ are separated by the gaps $\Delta\gtrsim E_{0}$ which do not have a hierarchical structure. Inset: probability density  of local level spacings.  \label{Fig:minibands} }
\end{figure}
Eq.(\ref{Ds1}) applies also in the case where $P(\Delta)$ is given by Eq.(\ref{eq:P(Delta)_power}) with $D_{s}<1$ for $\Delta<E_{0}$ and by
\begin{equation}
P(\Delta) \sim  \frac{C}{\Delta^{2}},
\end{equation}
for $\Delta>E_{0}$, where $E_{0}>E_{{\rm Th}}$ is some crossover scale that may depend on $N$, e.g. $E_{0}\sim N^{-z}$. In this case the normalization of $P(\Delta)$ (that is still determined by the small $\Delta\sim E_{{\rm Th}}$) requires the constant $C$ to be equal to:
\begin{equation}\label{C}
C\sim (E_{{\rm Th}})^{D_{s}}\,(E_{0})^{1-D_{s}}.
\end{equation}
Comparing Eqs.(\ref{Ds1}),(\ref{C}) with Eq.(\ref{Kw}) one concludes that  in the interval $E_{0}\lesssim \omega \lesssim N^{-\beta}\,(E_{0})^{1-D_{s}}$ the correlation function $K(\omega)$ acquires a "high-energy plateau" where it  depends on $\omega$ very slowly:
\begin{equation}\label{HE-plateau}
K(\omega)\sim \frac{(E_{{\rm Th}})^{-D_{s}}}{M\,(E_{0})^{1-D_{s}}}\,\frac{1}{\ln(\omega/E_{0})}\propto \frac{N^{\beta+z\mu}}{\ln(\omega/E_{0})}.
\end{equation}
The numerical results for $K(\omega)$ presented in the next section seem to indicate on existence of such a plateau. This result implies that the  frcatality of local energy spectrum with {\it hierarchy of mini-bands} exists in this model at small energy scales ${E_{{\rm Th}}}\lesssim \omega \lesssim E_{0}$, while the large gaps between "random Cantor sets" do not show a hierarchical structure (see Fig.\ref{Fig:minibands}).

\section{Numerical results for $K(\omega)$.} \label{numerics-Kw}

\begin{figure}[t!]
\begin{centering}
\includegraphics[width=8.20cm]{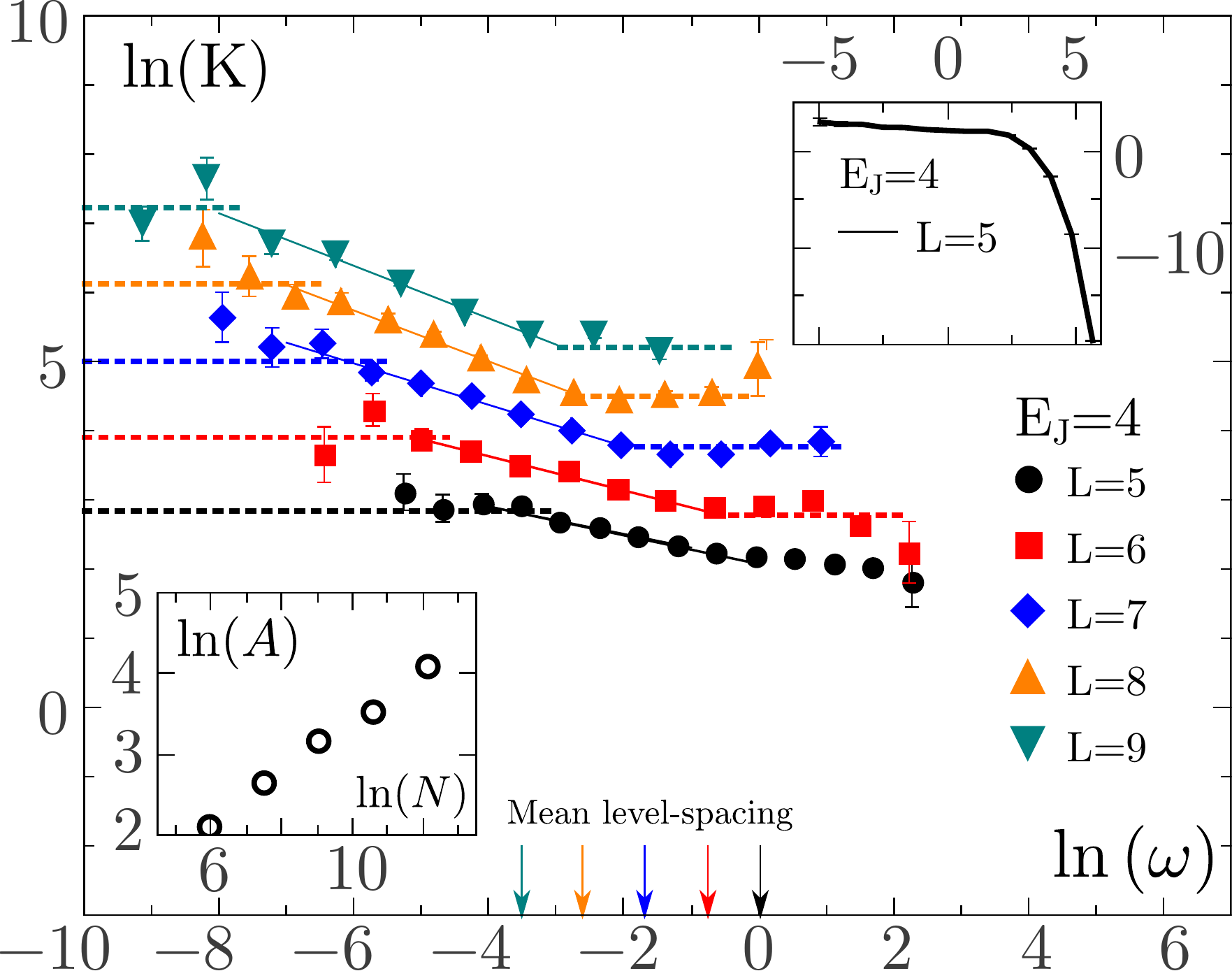}
\par\end{centering}
\caption{Logarithm of the LDoS correlation function, $\ln K_{E}(\omega)$, as a function
of the logarithm of the energy difference $\ln \omega$ in the multifractal
regime, $E_{J}=4$. Each data set corresponds to a different system
size $L$. The arrows indicate the many-body mean-level spacing, $\delta=(\rho(E)\,N)^{-1}$, for
each of the sizes. Partial diagonalization is used to compute $\ln(K_{E})$
with a few eigenstates at the reduced energy  $\bar{\epsilon}=0.1$ in
the main panel. The solid lines are fits $\ln K_{E}=\ln A-\mu\ln\omega$
that gives $\mu=0.20,\ 0.25,\ 0.30,0.36,\ 0.38$ for $L=5,6,7,8,9$
respectively. The dotted horizontal lines at low frequencies represents $\ln(N \, I_{2} /3)$,
where $I_{2}=\sum_{i}\overline{|\psi_{\alpha}(i)|^{4}}$. 
Intersection of the dotted and solid lines corresponds to $\omega=E_{{\rm Th}}$, see Eq.(\ref{Thouless-def}). 
The horizontal dotted lines at higher frequencies show an approximate "high-energy plateau". The frequency $\omega=E_{0}$  at the onset 
of this plateau corresponds approximately to the global mean level spacing $E_{0}\approx \delta\sim N^{-0.6}$.   Inserts: (top) the dependence
$\ln(K_{E}(\omega))$ in the whole range of $\omega$ obtained from
full diagonalization for $L=5$ and (bottom) the logarithm of the pre-factor $A$ in Eq.(\ref{def-beta-mu}) as a function
of $\ln(N)$.\label{fig4}}
\end{figure}

The results of the numerical computation of $K(\omega)$
are shown in Fig. \ref{fig4} for the intermediate value
of $E_{J}=4$.

The largest size displayed in this figure corresponds to Hilbert space
size $N\sim10^{6}$ and the statistics of $\sim10^{4}$ samples.

Two features are remarkable: $K_{E}(\omega)$ has a power law dependence
in a wide frequency interval $E_{Th}<\omega<E_{0}$ which is well described by Eq.(\ref{def-beta-mu}),
and the exponents of this power law are non-trivial ($\mu\approx0.4$,
$\beta\approx0.3$). Using the theory of Sec.\ref{sec:correlations} we may extract the fractal dimensions $D_{2}\approx 0.3$ and $D_{s}\approx 0.6$ in the Hilbert and energy space.
The observed power-law and the values of the critical exponents consistent with the theory is a strong argument in favor of the statement that for this choice of parameters of the model ($E_{J}=4$, $E_{C}=1$, $W=10$, $\bar{\epsilon}=0.1$) the system is in the non-ergodic, multifractal phase.
The fractal structure
of the local energy spectrum implies that in this regime the wavefunction
is first spread over a small cluster of close resonances, these resonances
are weakly entangled with another cluster further away to form a supercluster,
etc. to eventually form a large scale hierarchical structure similar to spin glasses.

The second feature is the "high-energy plateau" shown by horizontal dotted lines. It is remarkable that the onset of this plateau (or the upper cutoff of the power-law, Eq.(\ref{def-beta-mu})) is approximately equal to the global mean level spacing $E_{0}\approx \delta=(\rho(E)\,N)^{-1}$. This scale is much larger than the Thouless energy only because the calculations were done at $\bar{\epsilon}=0.1$ where the mean DoS $\rho(E)\sim 10^{-4}$ is very small. In agreement with the theory of the previous section this implies (see Fig.\ref{Fig:minibands}) that the hierarchical structure of gaps between mini-bands in the local energy spectrum exists only up to the scale coinciding with the global mean level spacing.

\section{ Scaling approach of the Many-Body Localization Transition}\label{sec:scaling}

\begin{figure}[t!]
\begin{centering}
\includegraphics[width=0.99\columnwidth]{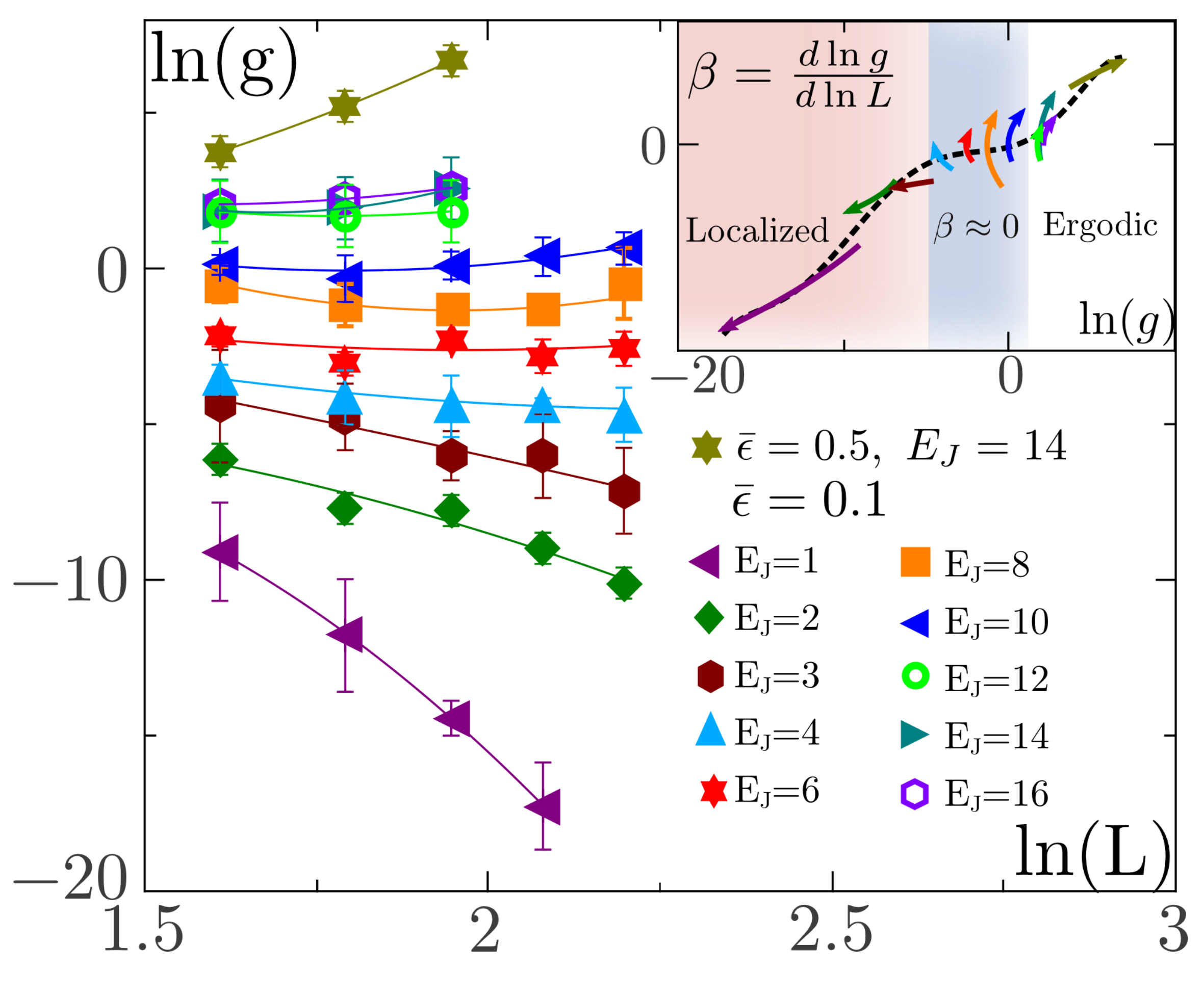}
\par\end{centering}
\caption{Logarithm of the Thouless conductance, $g=E_{{\rm Th}}/\delta$, as a
function of the dimension $N$ of the Hilbert space. The up-most curve in the
main panel has been computed in the middle of the band, $\bar{\epsilon}=0.5$.
All other curves correspond to the energy $\bar{\epsilon}=0.1$. The
insert shows the $\beta(g)=d\ln g/d\ln N$ function computed from these data. The
dotted line is a fit of all the $\beta$ functions for different sizes
to a 5th order polynomial. \label{fig5} }
\end{figure}

The data shown in Fig. \ref{fig4} and similar data
for different $E_{J}$ can be used to compute the Thouless energy
defined by Eq.(\ref{Thouless-def}).   In order to avoid direct
computation of $K(\omega)$ at low frequencies   we  used the result
for the participation ratio $I_{2}=\overline{\sum_{i}|\psi_{\alpha}(i)|^{4}}$.
As explained in Sec.\ref{sec:correlations}, one expects that $K(\omega\rightarrow0)\rightarrow c\,N\,I_{2}$ where
$c\sim1$. For Gaussian random matrix $c=1/3$, we shall use this
value because it agrees very well with the results of the computation
at small sizes $L$ for which we were able to compute $K(\omega)$
for small $\omega$ directly. Note that direct computation of $K(\omega)$
at $\omega\rightarrow0$ is very difficult due to a small number of
states with small energy differences. Combining the asymptotic at
very small frequencies (shown by dashed horizontal lines in Fig.\ref{fig4}) with the power law dependence at $E_{{\rm Th}}<\omega< E_{0}$ (shown by tilted solid line)
we determined the crossover frequency $E_{{\rm Th}}$ as the frequency where these two lines intersect. Then we obtain the many-body Thouless conductance $g=E_{{\rm Th}}/\delta$ shown in Fig. \ref{fig5}.

The size dependence of the Thouless conductance displays three distinct
regimes. For $E_{J}\lesssim3$ it decreases exponentially with the
system size $L\approx \ln N/\ln 5$ (or as  power-law  with  the dimension $N$ of the Hilbert space), similar to localized regime in conventional single-particle
theory. However, the slope   of the power-law $N$-dependence decreases as $E_{J}$ increases. The value of $E_{J}$ where the slope vanishes coincides well with the critical value $E_{J}^{(cr)}\approx3.5$ of the full many-body localization found from the level statistics (see section \ref{sec:statistic}).

In the interval $10\gtrsim E_{J}\gtrsim 4$ the Thouless conductance stays almost constant as $N$ changes by two orders of magnitude.
Notice that
the decrease of $g(L)$ disappears  when $g\sim 10^{-2}$ is very small and it stays $L$-independent in a wide interval of $E_{J}$ where $g(E_{J})$ changes by three orders of magnitude from $\sim 10^{-2}$ to $\sim 10$.

Only at  $E_{J}\gtrsim 14$  an exponential
increase with the system size $L$ is observed, signaling the appearance
of a conventional ergodic state. This increase is still within the error bars for
$\bar{\epsilon}=0.1$ but it becomes unquestionable in the band center $\bar{\epsilon}=0.5$.

These three regimes are shown in the inset of Fig.\ref{fig5} where $d\ln g/d \ln N$  is presented as a function of $g$. The appearance of a wide interval of $g$ where $\beta(g)=d\ln g/d \ln N$ is nearly constant is a remarkable feature of our model which allows to make a conclusion about existence of a non-ergodic extended phase
of a {\it bad metal}, or {\it critical metal}, in the Josephson Junction Array model under consideration.

This behavior is in a sharp contrast
with that for three-dimensional localization, in which case the conductance varies
exponentially with the system size $L$ for small $g$, is a   power law in
the system size for large $g$, and only in the critical point of the localization transition, where $g\sim 1$, it is $L$-independent. This difference is due to the fact that
in three dimensions the probability to find a resonance  site within
the energy interval $\Delta E$  at a distance $R$  increases as a
power of the distance   whereas the tunneling amplitude
decreases exponentially with $R$.
Since conductance
at size $L$ is proportional to the tunneling amplitude at this size,
at small conductance $g\ll g_{c}\sim1$ the virtual processes in which
the particle hops to the state close in energy in order to cross the
sample of size $L$ become improbable. In the many-body localization, the Hilbert space
has a local tree structure in which the probability to find a resonance site at large distance increases exponentially and can compensate for exponential decrease of the tunneling amplitude.

\section{Fractal dimensions of wave functions in the Hilbert space.}\label{sec:fractaldimensions}

Existence of the intermediate non-ergodic phase for $4\lesssim E_{J}\lesssim 10$
is in a full agreement with the analysis of the wavefunction moments,
defined by
\begin{equation}\label{I_q}
I_{q}=\sum_{i}\overline{|\psi(i)|^{2q}}.
\end{equation}
In a multifractal phase $I_{q}\propto N^{-D_{q}\,(q-1)}$. In a generic case the fractal dimensions:
\begin{figure}[t!]
\begin{centering}
\includegraphics[width=0.9\columnwidth]{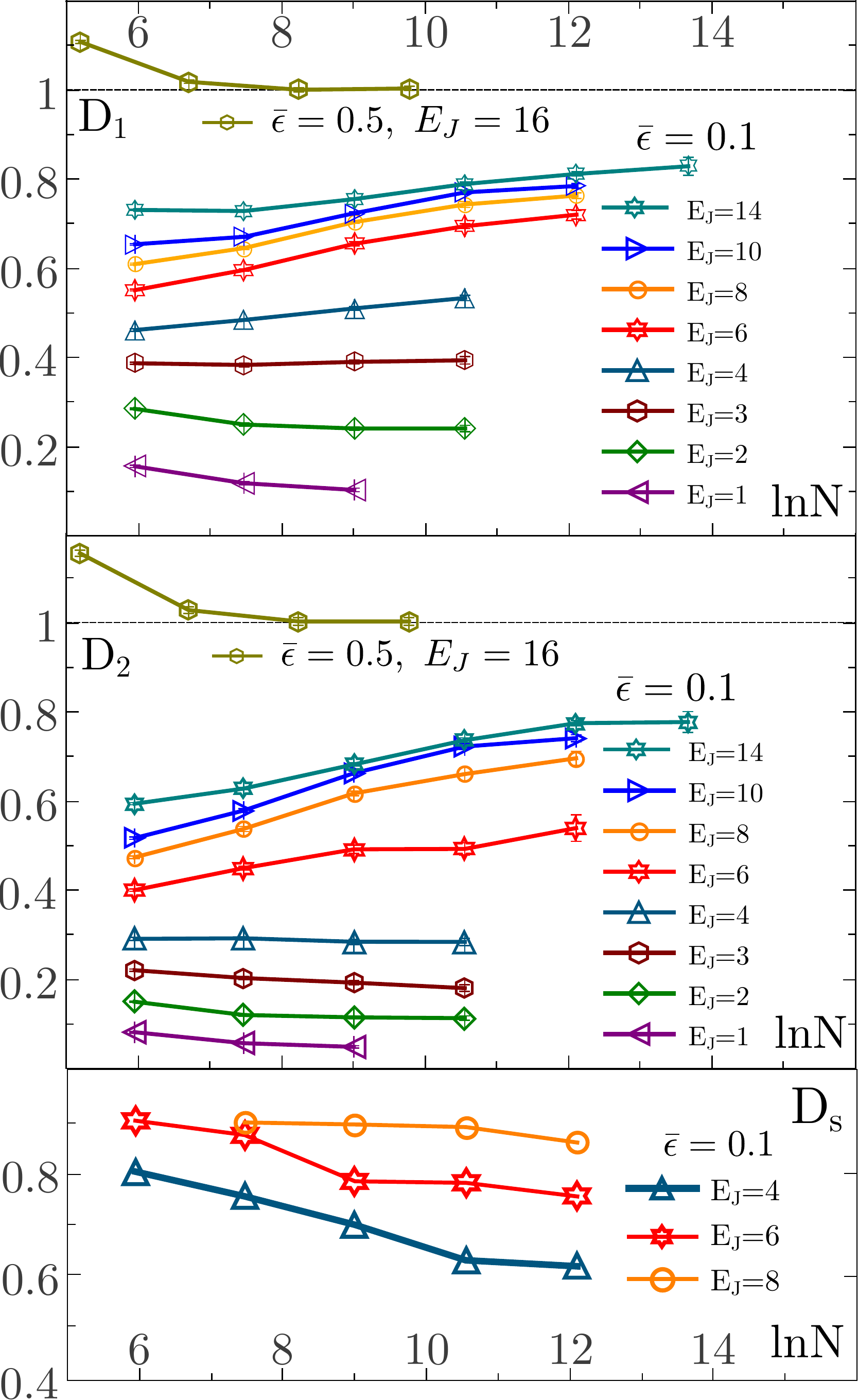}
\par\end{centering}
\caption{Fractal dimensions ($D_{1},$ $D_{2}$ and $D_{S}$) as a function
of logarithm of dimension $N$ of the Hilbert space. Each data set corresponds to a given
value of Josephson coupling $E_{J}$. Energy density is $\bar{\epsilon}=0.1$
for all points except one which corresponds to the middle of the spectrum
$\bar{\epsilon}=0.5$. \label{fig6}}
\end{figure}

\begin{align}
D_{q} & =-\frac{1}{q-1}\,\left(\frac{d\ln I_{q}}{d\ln N}\right),\label{Eq:FD1}
\end{align}
depend on the order of the moment $q$. The most popular for applications are $D_{q}$
with  $q=2$ and $q=1$ which  is understood as the limit of $D_{q}$ as $q\rightarrow1$.

We  computed dimensions $D_{1}$ and $D_{2}$ by employing the discrete, finite-size
version of Eq.\ (\ref{Eq:FD1})   in which the data for sizes $L+1$ and $L-1$  was used to compute $D_{q}(L)$.
In Fig. \ref{fig6},
$D_{q}$ is shown as a function of $\ln N$ (remember that $N\sim 5^{L}$ is the dimension of the Hilbert space) for different Josephson coupling ranging from $E_{J}=1$
to $E_{J}=14$.
For very small $E_{J}\leq 2$, the fractal dimensions definitely decrease
with the system size increasing and are likely to tend to zero in the limit $N\rightarrow\infty$. This is the signature of the insulating phase.
At $E_{J}=3$ the behavior is marginal with a very slow variations within the error bars. Starting from $E_{J}=4$ the increase 
of fractal dimensions becomes progressively more pronounced, signalling on the delocalized phase. 
Unfortunately, the evolution is too slow to converge to $N=\infty$ limiting behavior, even at a system size $N\sim 2\times 10^{5}$. 
This is a typical problem for systems with exponential proliferation of sites which was earlier encountered 
on a Bethe lattice and Random Regular Graph \cite{Al2016, Tikhonov2016b}. 
It does not allow to determine with absolute certainty the value of $D_{q}(\ln N)$ in the thermodynamical limit.
In any case, a clear signature of multifractality is the fact 
that $D_1$ is significantly larger than $D_2$ for a wide range of parameters.
Together with the size-independence of the  Thouless conductance in the interval
$4\lesssim E_{J}\lesssim 8$ established in Sec.\ref{sec:scaling} this result sets the lower bound  $E_{J}\gtrsim 8$ for the ergodic phase   at $\bar{\epsilon}=0.1$.
The result in the center of the band $\bar{\epsilon}=0.5$ 
for $E_{J}=14$ is also consistent with the conclusion of sec.\ref{sec:scaling} that   this choice of parameters clearly corresponds to the ergodic phase.

We conclude that both the results on the scaling of Thouless energy and the results of the size dependence of $D_{1}$ and $D_{2}$ show 
consistently that at $\bar{\epsilon}=0.1$ in the interval $4\lesssim E_{J}\lesssim 8$ the non-ergodic extended, multifractal  phase is present in our model.
We also note that the multifractality is strong in the range $E_{J}=4$ to $E_{J}=6$, as $D_{1}$ is significantly larger
than $D_{2}$.

\section{Relationship between critical exponents \label{sec:relationship}}
A simple theory of fractality of local energy spectrum presented in Sec.\ref{sec:correlations} suggests a relationship between the   exponents
$D_{s}=1-\mu$ and $\beta$ describing fractality of the local energy spectrum  and the fractal dimension $D_{2}$ of wave functions in the Hilbert space.
Combining Eqs.(\ref{betamu1}),(\ref{betamu2}) one obtains:
\begin{equation}\label{scaling-rel}
D_{s}=1-\mu=\beta+D_{2}.
\end{equation}
In the lower panel of Fig.\ref{fig6} we present the data for the fractal dimension $D_{s}=1-\mu$ of local energy spectrum determined from the power-law behavior of $K_{E}(\omega)$. Qualitatively its behavior as a function of $E_{J}$
confirms expectation that multifractality in the Hilbert space and in the energy space
are related and both becomes stronger (smaller fractal dimensions) as $E_{J}$ decreases. More quantitative results are presented in Table I.
\begin{table}[h!]
\centering \caption{Exponents $\mu, \beta, D_{2}$ and a test of the scaling relationship Eq.(\ref{scaling-rel}) between them. The exponents $\mu$ and $D_{2}$ were determined at the largest size of the system for which numerics were available.  }
\label{tab:table1} %
 \begin{tabular}{|c||c|c|c|}
\hline
$E_{J}$    &4    & 6     & 8\tabularnewline
\hline
$\beta$ & 0.32  & 0.32  & 0.29    \tabularnewline
\hline
$D_{2}$ & 0.23  & 0.53  & 0.68    \tabularnewline
\hline
$\beta+D_{2}$& 0.55  & 0.85  & 0.97   \tabularnewline
\hline
$1-\mu$ & 0.62  & 0.76  & 0.87  \tabularnewline
\hline
\end{tabular}
\end{table} 

Given  poor accuracy of $D_{2}$ and $\mu$ which significantly vary with increasing $\ln N$, the fulfilment of the scaling relationship Eq.(\ref{scaling-rel}) is very satisfactory.

This is a strong argument in favor of the theory described in Sec.\ref{sec:correlations} which is entirely based on the assumption of multifractality as the simplest form of non-ergodicity in the delocalized phase.

\section{Statistic of eigenenergies}\label{sec:statistic}
Finally we present the results for the so called $r$-statistic, which is the mean ratio of consecutive level spacings $\delta_{n}=E_{i+1}-E_{i}$ in the global spectrum: $$r=\overline{\min(\delta_{i},\delta_{i+1})/\max(\delta_{i},\delta_{i+1})}.$$
It is a popular measure 
to distinguish between  the MBL localized and extended phases \cite{Oganesyan2007,Cuevas2012}.
For the Wigner-Dyson distribution which corresponds to extreme delocalized, ergodic regime $\left\langle r\right\rangle =0.536$,  while for the Poisson distribution  expected in the localized phases $\left\langle r\right\rangle =0.386$  
\ \cite{Atas2013}. The crossing point of the curves $r_{N}(E_{J})$ for different   sizes $N$ marks the many-body localization transition.

\begin{figure}[t!]
\begin{centering}
\includegraphics[width=0.96\columnwidth]{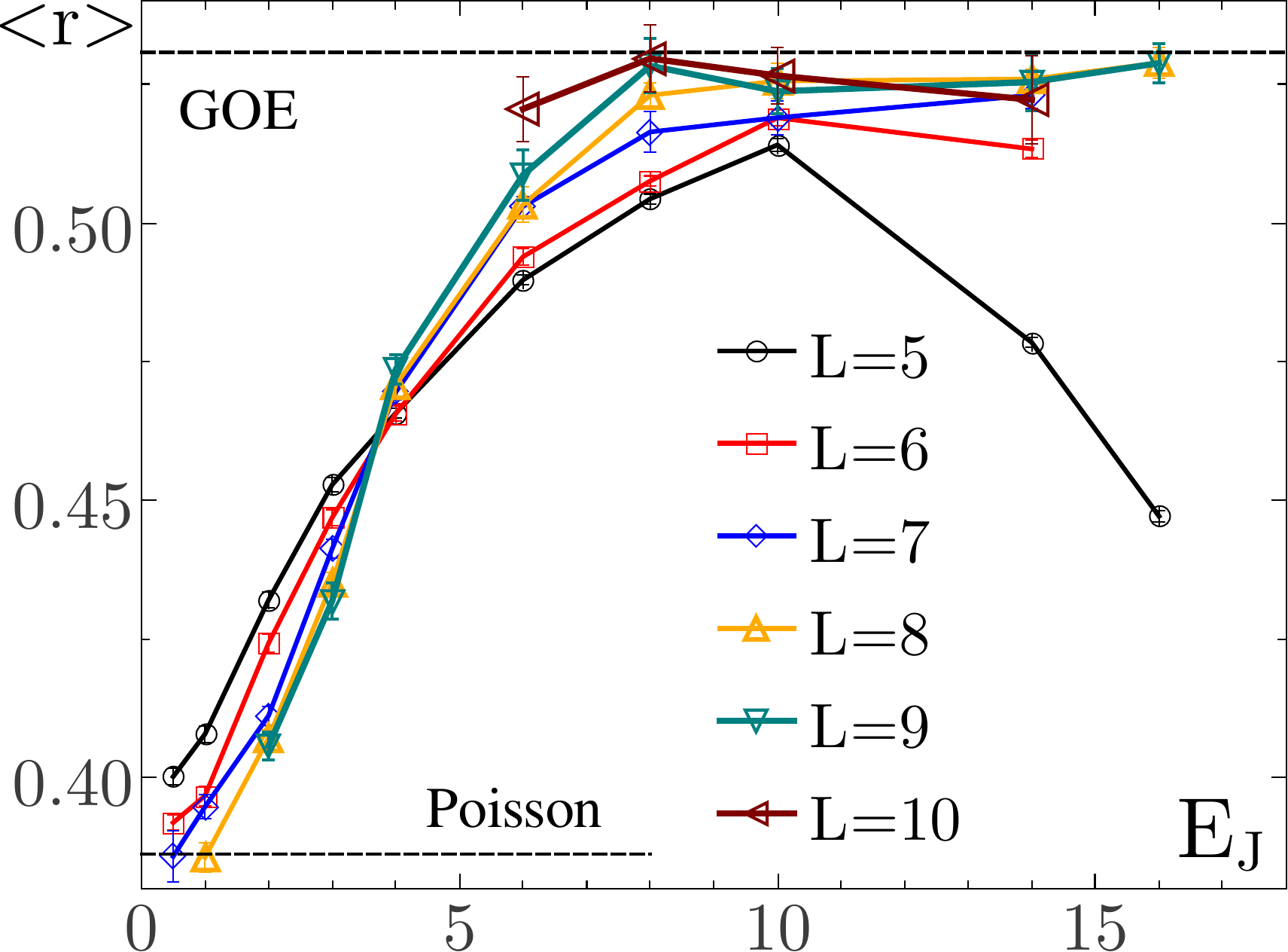}
\par\end{centering}
\caption{Ratio of minimum and maximum consecutive  global level spacings,
$r=\overline{\min(\delta_{i},\delta_{i+1})/\max(\delta_{i},\delta_{i+1})}$,
for eigenenergies at energy $\bar{\epsilon}=0.1$. Sizes run from
$L=5$ to $L=10$, as indicated in the legend. The horizontal lines
represent the value of $r$ for the Gaussian
Orthogonal ensemble $r_{{\rm WD}}=0.536$ and for
the Poisson level statistics $r_{{\rm P}} =0.386$.\label{fig3}}
\end{figure}

Fig.\ \ref{fig3}, presents $  r_{N}(E_{J}) $ for the levels at energy
$\bar{\epsilon}=0.1$ as a function of $E_{J}$ for several sizes $L=4,5\dots10$.
One observes an apparent crossing point at $E_{J}\approx 3.5$. The spread of curves
at $E_{J}>3.5$ clearly indicates to the delocalized phase, while that for $E_{J}<3.5$
shows an insulating behavior only for sizes $L=5,6,7$. For larger sizes the curves are almost coinciding which leaves a possibility that the MBL localization transition is somewhat lower than $E_{J}=3.5$.

An important feature of Fig.\ref{fig3} is that the apparent crossing happens  approximately half way from the Poisson to the Wigner-Dyson limits. This is contrary an expectation for the Anderson transition on the hierarchical networks such as  Bethe lattice or Random Regular Graph, where the transition point is very close to the Poisson limit. Given a very steep descent of the curves for large $L=7,8,9$ at small $E_{J}$ and the fact that all these curves are almost coinciding, one may expect that
the true crossing point corresponds to a value of $r$ much closer to the Poisson limit
then that for the apparent crossing point and the critical value for $E_{J}$ is in the interval $2<E_{J}<3$, in agreement with the results of the previous section.

\section{Conclusion.} Our results confirm  existence of the
multifractal regime {\it at least} in the interval $4\lesssim E_{J}\lesssim10$ for
the model Eq. (\ref{Eq:Ham}) of the Josephson junction chain. This conclusion is 
reached by comparing the two sets of data: the correlation function $K(\omega)$ of the local density of states which encodes the property of the local energy spectrum, and the eigenfunction moments
$I_{q}$ which contain information on multifractality in the Hilbert space. 
In this regime both the local
energy spectrum and the eigenfunction structure in the Hilbert space are fractal, see Fig. \ref{fig6}.
The scaling behavior is characterized by the size-independent many-body Thouless
 conductance that varies by orders of magnitude as a function
of $E_{J}$. This finding is hardly compatible with the single parameter
scaling   because it leads to a very abnormal $\beta=d\ln g/dL$ function
shown in Fig. \ref{fig5} (see also Ref.\cite{Garcia2016_2} for violation of single-parameter scaling
on Random Regular Graphs).
We would like to mention
the soluble 1D model \cite{Kovrizhin2017} that exhibits similar behavior due to exact conservation
laws. Similar to Josephson junction
chain, the number of states per site in this model is larger than
2 which indicates that absence of non-ergodic regime claimed in some
previous works\ \cite{Silva2014,Serbyn2016,Serbyn2016_2} might be
due to the choice of the model with only 2 states per site in 1D chain.

The appearance of a peculiar
regime in which $\beta(g)$ function is approximately zero in a wide range of parameters
is a clear evidence
for the new genuine phase, a ``bad'' metal. Physically, in this
phase one should observe dissipation and transport but the dynamics is slow
and thermodynamic equilibrium is never reached. One experimental evidence
for such state is a strongly enhanced noise and the violation of FDT.
Another evidence is the fractal nature of the local spectra.
The observation of the
fractal structure of the local spectra is of principal importance. It opens up new
direction of investigation of the "bad metal" phase by various spectral techniques.

\textbf{\emph{Acknowledgment. }}This work was supported by by ARO
grant W911NF-13-1-0431 and Russian Science Foundation 14-42-00044.
M. P. acknowledge support from Juan de la Cierva IJCI-2015-23260,
MINECO/FEDER Project FIS2015-70856-P, CAM PRICYT Project QUITEMAD+
S2013/ICE-2801 and Proyecto de la Fundacion Seneca 19907/GERM/15.

 \bibliographystyle{apsrev4-1}
\bibliography{MBLandNonequilibrium}

\end{document}